\documentclass{article}
\usepackage{graphicx} 
\usepackage{authblk}
\usepackage[numbers,square,sort&compress]{natbib}

\usepackage[utf8]{inputenc}
\usepackage[T1]{fontenc}
\usepackage{amsmath}
\usepackage{amssymb}
\usepackage{booktabs}
\usepackage{algorithm}
\usepackage{algpseudocode}
\usepackage{xurl} 

\usepackage{tikz}
\usetikzlibrary{positioning,fit,backgrounds,calc,arrows.meta,shapes.geometric}
\tikzstyle{startstop} = [ellipse, draw, minimum width=2.0cm, minimum height=1cm]
\tikzstyle{decision} = [diamond, draw, aspect=2, minimum height=1.5cm]
\tikzstyle{process} = [rectangle, draw, minimum width=2.8cm, minimum height=1cm]
\tikzstyle{arrow} = [->, thick]
\tikzstyle{box} = [draw, dotted, inner sep=0.5cm]
\usepackage{varwidth}

\title{Staging by the Book: Automatic Sleep Stage Classification Using Scoring Rules}

\author[1,2]{Emil Hardarson}
\author[2,3]{Konstantin Popov}
\author[2]{Sigridur Sigurdardottir}
\author[1,2]{Anna Sigridur Islind}
\author[1,2,3]{Erna Sif Arnardóttir}
\author[4,1]{María Óskarsdóttir}

\affil[1]{Department of Computer Science, Reykjavik University, Reykjavik, Iceland}
\affil[2]{Reykjavik University Sleep Institute, Reykjavik University, Reykjavik, Iceland}
\affil[3]{Department of Engineering, Reykjavik University, Reykjavik, Iceland}
\affil[4]{School of Mathematical Sciences, University of Southampton, Southampton, United Kingdom}

\affil[ ]{\textit{Corresponding author: Emil Hardarson, \texttt{emilh@ru.is}}}

\date{}

\begin{document}
\maketitle

\renewcommand\thefootnote{}
\footnotetext{\textbf{Abbreviations:} AI: Artificial Intelligence, ML: Machine Learning, PSG: Polysomnography, EEG: Electroecephalography, EOG: Electrooculography, EMG: Electromyography, AASM: American Academy of Sleep Medicine, R\&K: Rechtschaffen and Kales, LAMF: Low amplitude mixed frequency, SWA: Slow wave activity, LLM: Large language model}

\begin{abstract}
Automated sleep staging is commonly approached as a supervised machine learning problem, with deep learning methods dominating recent research. While machine learning models achieve near-human level agreement with human-scored reference sleep stages, their decisions are typically opaque and not designed to follow clinical scoring rules. We propose a transparent alternative: a deterministic, rule-based sleep staging method that explicitly operationalizes the American Academy of Sleep Medicine's (AASM) scoring logic as executable code, coupled with epoch-level natural-language justifications derived from an explanation trace. We evaluate the approach on 50 polysomnography recordings with a 10-scorer majority-vote consensus as reference. Across all recordings, the method agreed with the majority-vote reference in 60.5\% of epochs ($\kappa=0.42$), with substantially higher agreement on a dataset used during development (77.1\%, $\kappa=0.61$). Agreement with the reference was highest for sleep stage N2 (recall 83.5\%) and moderate for sleep stage R (recall 68.7\%), while Wake and N1 recall were low. Despite lower agreement with the reference than contemporary deep learning models, the method provides deterministic decisions and natural language explanations aligned with AASM scoring rules, making it a complementary tool for auditing, debugging, and governing deep learning-based sleep staging.
\end{abstract}

\textbf{Keywords:} sleep staging, rule-based AI, explainable AI, polysomnography

\section{Introduction}\label{sec:introduction}

Many decisions in clinical practice are governed by written rules and guidelines. 
When such decisions are automated, the goal is not only to produce an accurate output, but also to preserve the logic by which the decision is supposed to be made under the rules \citep{amann_explainability_2020,labkoff_toward_2024}.
Automation in clinical decision-making is increasingly performed with supervised Machine Learning (ML) algorithms, which learn statistical patterns that produce high agreement with human labels, without being explicitly constrained to the rules governing the decision \citep{rudin_stop_2019,hulsen_explainable_2023,amann_explainability_2020}.

Sleep stage scoring is an example of such a rule-governed clinical decision-making workflow. It is a foundational step in sleep research and clinical practice, serving as the basis for diagnosing sleep disorders and understanding the physiology of sleep.
The gold standard for measuring sleep is a sleep study called polysomnography (PSG), which involves the overnight recording of multiple physiological signals, such as electroencephalography (EEG), electrooculography (EOG), and chin electromyography (EMG) \citep{troester_aasm_2023}.

Clinical sleep stage scoring guidelines, such as those published by the American Academy of Sleep Medicine (AASM), specify that sleep should be staged in consecutive 30-second epochs. Each epoch is classified as one of five sleep stages: Wake, Rapid Eye Movement (REM) sleep (sleep stage R), or Non-REM sleep 1, 2, or 3 (sleep stages N1, N2, or N3). Sleep stage scoring is performed manually by a human sleep expert, in a procedure that typically takes between 1 and 3 hours for each full night recording \citep{fischer_standard_2012,malhotra_performance_2013}.
Manual sleep stage scoring is subject to inter-scorer variability, and agreement differs between sleep stages.
Multi-scorer studies have found relatively high agreement for sleep stages Wake, N2, and R, but lower agreement for sleep stages N1 and N3 \citep{rosenberg_american_2013,nikkonen_multicentre_2024,lee_interrater_2022}.
Complete agreement across expert scorers occurs in only a minority of epochs \citep{bakker_scoring_2023,anderer_overview_2023}.
The growing number of sleep studies \citep{braun_systematic_2024} and the labor-intensive nature of manual sleep stage scoring have motivated the development of practical and reliable automatic sleep staging algorithms \citep{fiorillo_automated_2019}.

In recent years, supervised ML, and specifically Deep Learning (DL), has become the dominant methodology for automated sleep staging \citep{alsolai_systematic_2022,zhang_eeg-based_2018,alattar_artificial_2024,gaiduk_current_2023}.
ML models trained on large datasets of human-labeled sleep recordings have achieved performance comparable to human inter-scorer agreement \citep{perslev_u-sleep_2021,vallat_open-source_2021}.
The ML models do this by learning to identify statistical patterns within the signals \citep{supratak_deepsleepnet_2017,phan_l-seqsleepnet_2023,hardarson_data-local_2026}. However, this training procedure is typically not guaranteed to align with the logic of clinical scoring rules \citep{stanley_future_2023}.
The reasoning behind the models' predictions is difficult to interpret and explain, a phenomenon termed the ``black-box'' problem \citep{hulsen_explainable_2023}.

This opacity of sleep staging ML model reasoning presents a barrier to their clinical adoption \citep{stanley_future_2023}. When a human sleep expert scores a sleep study, they can justify their decision by pointing to features defined in clinical guidelines, such as sleep spindles or alpha rhythm \cite{hardarson_human-ai_2024}. Although ML models may learn to recognize such patterns, they do so by mapping high-dimensional data through millions of internal parameters, rather than through explicit logical steps.

Sleep stage scoring is, by design, a rule-based decision process whose underlying logic is already specified. The AASM scoring manual provides a deterministic set of rules, so an algorithm does not, in principle, need to infer this logic from training data.
Despite this, there is a notable gap in contemporary literature regarding methods that explicitly use the scoring logic described in clinical guidelines.

Historically, automated sleep staging did not begin with ML. Early methods attempted to translate human scoring practice into algorithms, first under the Rechtschaffen and Kales (R\&K) criteria \citep{rechtschaffen_manual_1968} and later under the first version of the AASM scoring manual \citep{iber_aasm_2007,hasan_past_1996,penzel_computer_2000}.
Earlier rule-based methods, such as that of \cite{liang_rule-based_2012}, have shown that sleep scoring logic could be operationalized with high agreement to human consensus \citep{penzel_computer_2000,anderer_e-health_2005,anderer_computer-assisted_2010,boira_validation_2025}.
The field of automated sleep staging later shifted towards ML and DL approaches because of their strong sleep staging accuracy \citep{fiorillo_automated_2019,faust_review_2019}, but rule-based approaches have not disappeared \citep{gunnarsdottir_novel_2020}.

A practical challenge in operationalizing sleep scoring rules is that the sleep stage label depends on intermediate signal features described in the scoring manual. 
For example, scoring decisions may depend on whether an epoch contains sleep spindles, K-complexes, slow wave activity, rapid eye movements, alpha rhythm, low-amplitude mixed-frequency activity, or changes in chin EMG tone. 
In this paper, we refer to these signal segments that are relevant to sleep stage scoring decisions as \textit{micro-annotations}.
Earlier computer-assisted and rule-based sleep staging systems have implicitly operationalized parts of clinical scoring logic, for example through expert-defined features or rules about sleep stage transitions \citep{penzel_computer_2000,anderer_e-health_2005,anderer_computer-assisted_2010,liang_rule-based_2012}. 
However, most contemporary automatic sleep staging systems report sleep stage labels or probabilities, rather than exposing the rule-application process that connects signal features to scoring decisions.

Beyond assigning a sleep stage label, an automated scoring method can also support interpretation by showing how the decision was reached. 
In sleep staging, this requires connecting signal features, scoring criteria, and the final stage assignment in a form that is familiar to human scorers \citep{hardarson_human-ai_2024}. 
Few automatic sleep staging systems provide epoch-level natural-language justifications that trace the decision through the scoring rules.

In this paper, we build on the rule-based tradition of automatic sleep staging by making the operationalization explicit: we translate the logic of the AASM scoring manual into executable code and record the reasoning process behind each sleep stage assignment. 
The result is a deterministic and inspectable method in which sleep stage assignments can be traced to the scoring rules and signal annotations used by the algorithm.

In this paper, we ask: To what extent can the logic of the AASM sleep stage scoring manual be operationalized as deterministic, inspectable code, and what limitations emerge when such a method is evaluated against multi-scorer human consensus?
We address this research question by developing a sequential, logic-driven sleep staging algorithm in Python that operationalizes the logic of the AASM scoring rules as executable code. In addition, we implement an explanation mechanism that records the rule-based reasoning behind each sleep stage assignment and renders this reasoning as natural-language justifications linked to scoring rules.

We evaluate the method using 50 full-night type II PSG recordings scored independently by 10 human sleep experts. The evaluation shows that the method achieves moderate agreement with the multi-scorer consensus, with the strongest performance for sleep stages N2 and R. A reviewer-based discrepancy assessment suggested that reviewed disagreements were primarily associated with incomplete or inaccurate micro-annotations, rather than the rule logic itself.
While previous rule-based sleep staging methods have used rules as a guideline or processing step, we present what is, to our knowledge, the first method that attempts to explicitly code the clinical logic of the AASM scoring manual into an algorithm \citep{yazdi_review_2024,alsolai_systematic_2022}.

\section{Method}\label{sec:method}

\subsection{Algorithm}

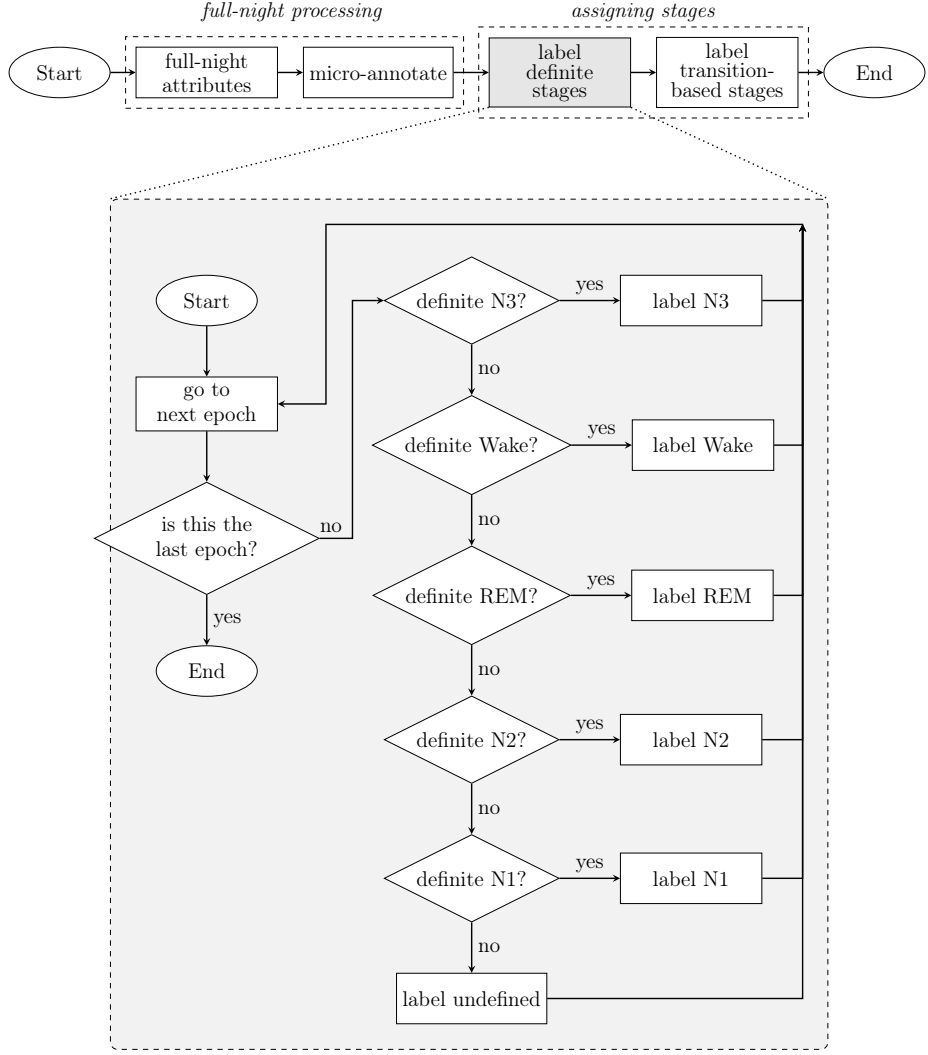
\begin{figure}[htbp]
\centering

\resizebox{\columnwidth}{!}{%
\begin{tikzpicture}[
    node distance=1.0cm and 1.2cm,
    every node/.style={font=\large, align=center},
    startstop/.style={ellipse, draw, minimum width=2.0cm, minimum height=1cm, fill=white},
    decision/.style={diamond, draw, aspect=2, minimum height=1.5cm, fill=white},
    process/.style={rectangle, draw, minimum width=2.8cm, minimum height=1cm, fill=white},
    box/.style={rectangle, draw, inner sep=0.5cm, dashed, rounded corners},
    arrow/.style={thick,->,>=stealth}
]

\node (start1) [startstop] {Start};
\node (attr) [process, right=0.5cm of start1] {\shortstack{full-night\\attributes}};
\node (micro) [process, right=0.5cm of attr] {micro-annotate};
\node (definite) [process, right=0.7cm of micro, fill=gray!20] {\shortstack{label\\definite\\stages}};
\node (transition) [process, right=0.5cm of definite] {\shortstack{label\\transition-\\based stages}};
\node (end1) [startstop, right=0.5cm of transition] {End};

\draw [arrow] (start1) -- (attr);
\draw [arrow] (attr) -- (micro);
\draw [arrow] (micro) -- (definite);
\draw [arrow] (definite) -- (transition);
\draw [arrow] (transition) -- (end1);

\node[draw, dashed, fit=(attr)(micro), inner sep=0.2cm] (box1) {};
\node at ($(attr)!0.5!(micro)+(0,1.2)$) {\large\textit{full-night processing}};

\node[draw, dashed, fit=(definite)(transition), inner sep=0.2cm] (topbox) {};
\node at ($(definite)!0.5!(transition)+(0,1.2)$) {\large\textit{assigning stages}};

\node (start2) [startstop, below=3.5cm of attr] {Start};
\node (nextepoch) [process, below=of start2] {go to \\next epoch};
\node (last) [decision, below=of nextepoch] {is this the\\last epoch?};
\node (end2) [startstop, below=of last] {End};

\node (defn3) [decision, right=2.5cm of start2] {definite N3?};
\node (labeln3a) [process, right=of defn3] {label N3};

\node (defwake) [decision, below=of defn3] {definite Wake?};
\node (labelwakea) [process, right=of defwake] {label Wake};

\node (defrem) [decision, below=of defwake] {definite REM?};
\node (labelrema) [process, right=of defrem] {label REM};

\node (defn2) [decision, below=of defrem] {definite N2?};
\node (labeln2a) [process, right=of defn2] {label N2};

\node (defn1) [decision, below=of defn2] {definite N1?};
\node (labeln1a) [process, right=of defn1] {label N1};

\node (labelundef) [process, below=of defn1] {label undefined};

\coordinate (toproute) at ($(labeln3a.north) + (0, 1.0cm)$);
\coordinate (farright) at ($(labeln3a.east) + (0.8cm, 0)$);

\coordinate (aisleDown) at ($(start2)!0.45!(defn3)$); 
\coordinate (aisleUp)   at ($(start2)!0.55!(defn3)$); 

\begin{pgfonlayer}{background}
    \node[box, fill=gray!10, fit={
        (start2) (nextepoch) (end2) 
        (labeln3a) (labelundef) (farright) (toproute) (aisleDown) (aisleUp)
    }] (groupbox) {};
\end{pgfonlayer}

\draw [arrow] (start2) -- (nextepoch);
\draw [arrow] (nextepoch) -- (last);
\draw [arrow] (last) -- node[right] {yes} (end2);

\draw [arrow] (last.east) -- node[above, pos=0.35] {no} (aisleUp |- last.east) -- (aisleUp |- defn3.west) -- (defn3.west);

\draw [arrow] (defn3) -- node[above] {yes} (labeln3a);
\draw [arrow] (defn3) -- node[right] {no} (defwake);

\draw [arrow] (defwake) -- node[above] {yes} (labelwakea);
\draw [arrow] (defwake) -- node[right] {no} (defrem);

\draw [arrow] (defrem) -- node[above] {yes} (labelrema);
\draw [arrow] (defrem) -- node[right] {no} (defn2);

\draw [arrow] (defn2) -- node[above] {yes} (labeln2a);
\draw [arrow] (defn2) -- node[right] {no} (defn1);

\draw [arrow] (defn1) -- node[above] {yes} (labeln1a);
\draw [arrow] (defn1) -- node[right] {no} (labelundef);

\draw [arrow] (labeln3a.east) -- (farright |- labeln3a.east) -- (farright |- toproute) -- (aisleDown |- toproute) -- (aisleDown |- nextepoch.east) -- (nextepoch.east);
\draw [arrow] (labelwakea.east) -- (farright |- labelwakea.east) -- (farright |- toproute);
\draw [arrow] (labelrema.east) -- (farright |- labelrema.east) -- (farright |- toproute);
\draw [arrow] (labeln2a.east) -- (farright |- labeln2a.east) -- (farright |- toproute);
\draw [arrow] (labeln1a.east) -- (farright |- labeln1a.east) -- (farright |- toproute);
\draw [arrow] (labelundef.east) -- (farright |- labelundef.east) -- (farright |- toproute);

\draw[dotted, thick] (definite.south west) -- (groupbox.north west);
\draw[dotted, thick] (definite.south east) -- (groupbox.north east);

\end{tikzpicture}%
}

\caption{Schematic overview of the automatic sleep staging process. The top diagram shows the high-level steps from initial processing of full-night attributes and micro-annotating to the two-pass sleep staging process. Full-night attributes refer to recording-level information, such as alpha-rhythm status and chin electromyography baseline, that is needed before applying some epoch-level scoring rules. The labeling of definite sleep stages is shown in an expanded view, where each epoch is sequentially compared to criteria for each of the sleep stages. Abbreviations: N1, non-rapid eye movement sleep stage 1; N2, non-rapid eye movement sleep stage 2; N3, non-rapid eye movement sleep stage 3; REM, rapid eye movement sleep.}
\label{fig:combined-sleep-staging}
\end{figure}

We designed the automatic sleep staging algorithm to emulate the logic of the AASM scoring manual. The method, written in Python 3.12, determines sleep stages by processing PSG data in three steps as shown in Figure \ref{fig:combined-sleep-staging}. 

The process begins by loading the raw PSG data stored in European Data Format (EDF), a standard file format for physiological time-series recordings, using the MNE Python library \citep{larson_mne-python_2025}, dividing the signals into 30-second epochs, and checking signal integrity. After this, the sleep staging process begins. First, the method establishes attributes pertaining to the full night's recording. These \textit{full-night attributes} act as global variables that govern the applicability of certain scoring rules. For example, the boolean variable \texttt{generates\_alpha\_rhythm} is decided via a sliding window Welch periodogram analysis on occipital EEG channels, which estimates the spectral power of the signal over time and identifies whether alpha-band activity is present  \citep{troester_aasm_2023,welch_use_1967}. This variable distinguishes between the use of the AASM scoring manual's \textit{F2 rule} (sleep should begin when alpha activity is attenuated and replaced by a low-amplitude mixed-frequency EEG signal in patients who generate alpha rhythm) and \textit{F3 rule} (sleep starts at the earliest occurrence of background frequency slowing, vertex sharp waves, or slow eye movements). Another full-night attribute is the chin EMG baseline, which establishes global statistics for the chin EMG signal in order to define thresholds for high and low muscle tone, which is used when classifying sleep stage R.

Second, the algorithm identifies various physiological events described in the scoring rules. These micro-annotations are stored as tuples containing event labels, temporal boundaries, and channel derivations. The YASA Python library \citep{vallat_open-source_2021} was used to detect sleep spindles, rapid eye movements (REMs), and slow wave activity (SWA) on frontal and central EEG channels. Alpha rhythm and low amplitude mixed frequency (LAMF) segments were operationalized using spectral features estimated with Welch periodograms \citep{welch_use_1967}. Specifically, power spectral density (PSD) was estimated in sliding windows, and each window was characterized by the relative power in frequency bands relevant to the AASM scoring rules \cite{troester_aasm_2023}. Windows meeting the predefined alpha or LAMF criteria were marked as corresponding micro-annotations. Alpha rhythm was detected using 2.0-second windows advanced in 0.5-second steps, corresponding to 75\% overlap. A window was marked as alpha rhythm when power in the 8--12 Hz band accounted for at least 50\% of total spectral power. LAMF was defined heuristically as a contiguous low-amplitude EEG segment lasting at least 1.0 second, where low amplitude was defined as absolute signal amplitude below the channel-level mean absolute amplitude minus 0.01 standard deviations. Candidate low-amplitude segments were then retained as LAMF if power in the 4--7 Hz band accounted for at least 1\% of total spectral power.

The algorithm processes the annotated epoch in two sequential passes to assign sleep stages, as shown in Figure \ref{fig:combined-sleep-staging}. The first pass evaluates epochs according to \textit{definite} scoring rules in the same order of precedence as the AASM scoring rules: sleep stage N3, then wake, R, N2, and finally N1, stopping at the first satisfied rule. Sleep stage N3 is assigned if SWA covers more than 20\% of the epoch. Only if that criterion is not fulfilled does the method check for evidence of the next sleep stage. Wake is assigned if alpha rhythm or other markers such as eye blinks exceed 50\% of the epoch. If not, the epoch is labeled sleep stage R if it contains LAMF and REMs without K-complexes or spindles, provided that EMG tone is low. Sleep stage N2 is assigned if a K-complex unassociated with arousal or a spindle occurs in the first half of the current epoch or the last half of the previous one. Finally, sleep stage N1 is assigned using a simplified alpha-rhythm rule. In alpha-generating recordings, the method assigns N1 when the preceding epoch was Wake, at least 50\% of the current epoch consists of LAMF, and alpha rhythm is detected in either the current or preceding epoch.

Epochs that remain undefined after the first pass are re-evaluated in a second pass using \textit{transition rules}, which depend on the context of adjacent epochs. For example, the method continues to score sleep stage R if the previous epoch was sleep stage R and the current epoch exhibits LAMF and low EMG without intervening arousals or spindles.

Epochs that remain undefined after both passes can either inherit the sleep stage of the preceding epoch or be assigned a sleep stage based on the YASA ML sleep stage classifier, depending on the user's preference. We report results for the inheritance-only configuration. Algorithm \ref{alg:sleep_staging} shows the step-by-step procedure for translating physiological micro-annotations into sleep stage labels in broad terms.

\begin{algorithm}
\caption{Rule-Based Sleep Staging}\label{alg:sleep_staging}
\begin{algorithmic}
\Require Epochs $\{E_t\}_{t=1}^T$ with micro-annotations.
\Ensure Sleep stage labels $y_t$ and explanations $\mathcal{E}_t$.

\State \textbf{Initialize:} Set all labels $y_t \gets \text{undefined}$

\vspace{1mm}
\State \textbf{Pass 1: Definite Rules (Local Epoch Criteria)}
\For{each epoch $E_t$ from $1$ to $T$}
    \State \textbf{1. Stage N3 Check:} If Slow Waves $\ge 20\%$ of epoch $\implies y_t \gets \text{N3}$
    \State \textbf{2. Wake Check:} If Wake-signs (Alpha, Blinks, etc.) $> 50\%$ $\implies y_t \gets \text{Wake}$
    \State \textbf{3. Stage R Check:} If (LAMF + REMs + Low EMG) and No (Spindle/K-Complex) $\implies y_t \gets \text{REM}$
    \State \textbf{4. Stage N2 Check:} If Spindle or K-Complex in current 1st half or previous 2nd half $\implies y_t \gets \text{N2}$
    \State \textbf{5. Stage N1 Check:} If (Alpha attenuated) and (LAMF $> 50\%$) $\implies y_t \gets \text{N1}$
    \State \textbf{Log Reasoning:} Update $\mathcal{E}_t$ with triggered rule and reasons for rejecting other rules.
\EndFor

\vspace{1mm}
\State \textbf{Pass 2: Transition Rules (Temporal Context)}
\For{each epoch $E_t$ where $y_t$ is still \textit{undefined}}
    \State \textbf{REM Continuity:} If $y_{t-1} = \text{stage R}$ and features remain consistent $\implies y_t \gets \text{stage R}$
    \State \textbf{N2 Continuity:} If $y_{t-1} = \text{stage N2}$ and no evidence of stage change $\implies y_t \gets \text{stage N2}$
    \State \textbf{Inheritance:} If still undefined, inherit stage from $y_{t-1}$.
    \State \textbf{Log Reasoning:} Document the use of preceding epoch in $\mathcal{E}_t$.
\EndFor

\State \Return Sleep stage labels and explanations $\{(y_t, \mathcal{E}_t)\}_{t=1}^T$
\end{algorithmic}
\end{algorithm}

\subsection{Explanation generation}

One of the strengths of a rule-based sleep staging algorithm is the possibility of providing transparent justifications for every staging decision. The method does this by populating an \textit{explanation log} as it traverses the logical steps. The algorithm appends the result of every sleep stage evaluation to a list within the explanation log, which is implemented as a nested dictionary. Each entry in the list is a tuple containing the sleep stage being evaluated, a boolean indicating if the criteria were met, and a text string detailing the findings. This creates a complete trace of the step-by-step decision process.

The explanation log shows how the final sleep stage is reached by sequentially ruling out alternative candidate sleep stages until the final decision is reached.
Because the algorithm stops searching once a definite sleep stage is assigned, the logs naturally present a chronological account of what was tried, why certain sleep stages were rejected (e.g., "slow wave activity at 15\%, below the 20\% threshold"), and which rule ultimately triggered the label.

The explanation log is a technical data structure designed for programmatic tracking of operations. To facilitate the user's deeper understanding of the logical reasoning behind staging decisions, the explanation log is fed into a Large Language Model (LLM; specifically OpenAI GPT-4). The LLM's system prompt is constructed with the epoch's explanation log and the Python source code used for the classification and explanation log. This provides the LLM with the context necessary to interpret the staging decision and justify it by referencing specific events in the signals, as well as the AASM scoring rules.

\subsection{Dataset}
\label{sec:dataset}

We evaluated our method on the Sleep Revolution Multicenter Scoring dataset, which consists of 50 prospective full-night type II PSG recordings collected at Reykjavik University (February - June 2021) using a Nox A1 system \citep{nikkonen_multicentre_2024}. Each recording was independently scored by 10 human sleep experts from seven sleep centers (April - September 2021) in Europe and Australia following AASM v2.6 criteria, enabling both a majority-vote consensus hypnogram and a per-epoch agreement ratio (fraction of scorers assigning the most frequently scored sleep stage). 

The cohort is adult and mixed with respect to common sleep disorders (healthy, obstructive sleep apnea, insomnia, and restless legs syndrome symptoms) (58\% male, mean age 42.9$\pm$13.7 years, mean BMI 27.3$\pm$5.8 kg/m$^2$, and mean AHI 15.2$\pm$15.6 events/h) \citep{nikkonen_multicentre_2024}.
One PSG was used in the development of the method for manual iterative rule implementation and parameter tuning and was subsequently excluded from the primary evaluation. The remaining 49 PSGs form the test set.

\subsection{Evaluation}

The performance of the method was evaluated using the dataset, with the majority vote consensus hypnogram of the ten scorers as a reference.
We assessed agreement both in terms of the epoch-by-epoch sleep stage classification and sleep architecture metrics. 
We calculated overall staging accuracy and Cohen's kappa, as well as confusion matrices, to identify sleep-stage-specific patterns of disagreement.
Before calculating agreement and sleep-architecture metrics, each recording was cropped to an analysis period. Following \citet{nikkonen_multicentre_2024}, this period was defined as the interval from the first epoch to the last epoch in which at least one human scorer assigned a sleep stage other than Wake. Leading and trailing all-Wake periods were removed, while Wake epochs within this interval were retained. The same cropping procedure was applied to the development recording and to all test recordings.

To complement this performance evaluation, a reviewer with sleep scoring experience conducted a reviewer-based discrepancy assessment to characterize sources of disagreement between the method and the reference. This assessment targeted two categories: (1) Epochs characterized by high human agreement where the method did not match the consensus, and (2) epochs with low human agreement where the method matched the sleep stage of the majority-vote consensus. A selection of epochs from both categories was investigated to determine whether the disagreement was associated with inaccurate or incomplete micro-annotations, rule application, or another source. By examining epochs with high human disagreement, our analysis sought to determine if the method's classification could provide insight into the reasons for ambiguity in human scoring.

We randomly sampled 20 epochs from the PSG used for method development, which had been used during iterative rule implementation and parameter tuning. For each sampled epoch, the reviewer first assigned a sleep stage using the PSG signals and the AASM scoring manual, without knowing the sleep stage label from the human sleep experts and the sleep staging algorithm. Next, the reviewer was shown the human sleep expert consensus sleep stage label and the agreement ratio, followed by the sleep stage label predicted by the algorithm and the algorithm's explanation. The reviewer then answered three yes/no questions: (1) Were the micro-annotations used by the algorithm correct? (2) Given those micro-annotations, was the applied scoring rule correct? (3) Was the assignment of definite or transition sleep stage label correct? If the reviewer determined that micro-annotations were incorrect, they could answer ``not applicable'' to the other two if deemed necessary.

\section{Results}\label{sec:results}

\subsection{Output data}

\begin{figure*}[t]
\centerline{\includegraphics[width=\textwidth]{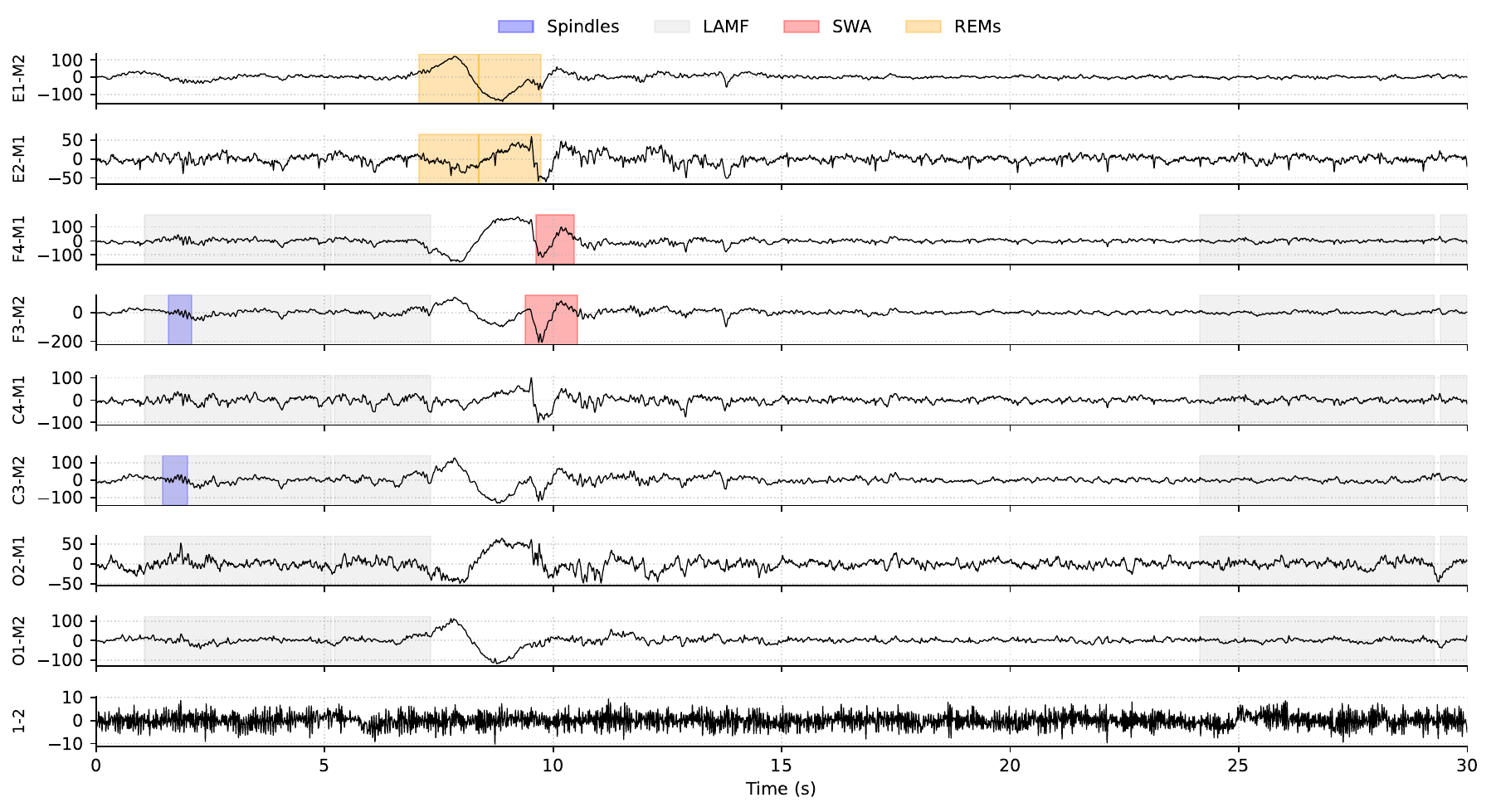}}
\caption{One epoch with micro-annotations. The displayed channels include electrooculography (EOG: E1-M2, E2-M1), electroencephalography (EEG: F4-M1, F3-M2, C4-M1, C3-M2, O2-M1, O1-M2), and chin electromyography (EMG: 1-2). Colored overlays indicate detected sleep spindles, low-amplitude mixed-frequency activity (LAMF), slow wave activity (SWA), and rapid eye movements (REMs). This epoch was classified as sleep stage N2. Abbreviations: EEG, electroencephalography; EOG, electrooculography; EMG, electromyography; LAMF, low-amplitude mixed-frequency activity; SWA, slow wave activity; REMs, rapid eye movements; N2, non-rapid eye movement sleep stage 2.}
\label{fig:example_epoch}
\end{figure*}

For each PSG recording, the method outputs a hypnogram consisting of epoch-by-epoch sleep stages, along with an epoch-by-epoch visualization of the micro-annotations. Additionally, the method produces the explanation log for every staging decision.
Figure \ref{fig:example_epoch} shows an example 
of an epoch which the algorithm classified as N2. The figure shows the signals used, along with the micro-annotations.

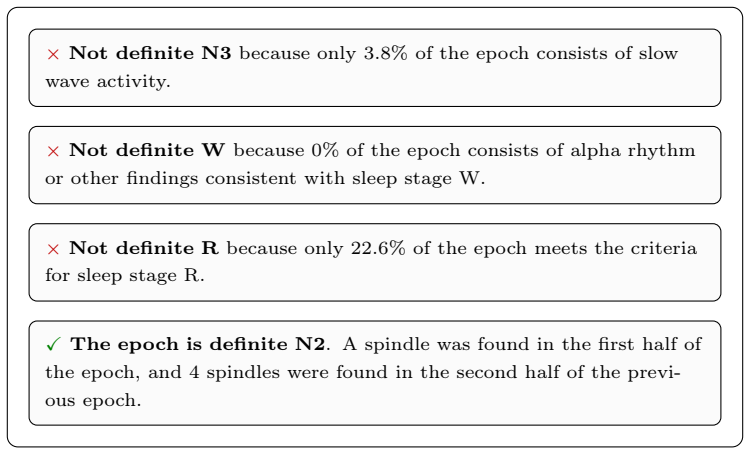
\begin{figure}[t]
\centering
\begin{tikzpicture}[
  font=\fontsize{7.0}{10.5}\selectfont,
  rejectbox/.style={
    draw,
    rounded corners=3pt,
    fill=gray!3,
    align=left,
    inner sep=6pt,
    text width=0.72\linewidth
  },
  finalbox/.style={
    draw,
    rounded corners=3pt,
    fill=gray!3,
    align=left,
    inner sep=6pt,
    text width=0.72\linewidth
  },
  markx/.style={
    text=red!75!black,
    font=\bfseries\large
  },
  marktick/.style={
    text=green!50!black,
    font=\bfseries\large
  },
  outerbox/.style={
    draw,
    rounded corners=4pt,
    inner sep=8pt
  }
]

\node[rejectbox] (r1)
{\textbf{\textcolor{red!75!black}{$ \times $} Not definite N3} because only 3.8\% of the epoch consists of slow wave activity.};

\node[rejectbox, below=2.5mm of r1] (r2)
{\textbf{\textcolor{red!75!black}{$ \times $} Not definite W} because 0\% of the epoch consists of alpha rhythm or other findings consistent with sleep stage W.};

\node[rejectbox, below=2.5mm of r2] (r3)
{\textbf{\textcolor{red!75!black}{$ \times $} Not definite R} because only 22.6\% of the epoch meets the criteria for sleep stage R.};

\node[finalbox, below=2.5mm of r3] (f1)
{\textbf{\textcolor{green!50!black}{$\checkmark$} The epoch is definite N2}. A spindle was found in the first half of the epoch, and 4 spindles were found in the second half of the previous epoch.};

\node[outerbox, fit=(r1)(r2)(r3)(f1)] {};

\end{tikzpicture}
\caption{Example of a sequential elimination trace produced by the rule-based sleep staging method. The explanation rules out incompatible sleep stages in sequence before stating the final sleep stage assignment and its supporting evidence. Abbreviations: W, Wake; R, rapid eye movement sleep; N2, non-rapid eye movement sleep stage 2; N3, non-rapid eye movement sleep stage 3.}
\label{fig:static_explanation}
\end{figure}

Along with inspecting the raw explanation log, the user has two ways to view explanations: \textit{static-} and \textit{interactive} explanations. In static explanations, the log is rendered as a list of elimination rules that outline why each candidate sleep stage was not assigned until the final rule triggers the selected label, in the order in which they were applied.
Figure \ref{fig:static_explanation} shows a static explanation generated by the algorithm for the same epoch as shown in Figure \ref{fig:example_epoch}.

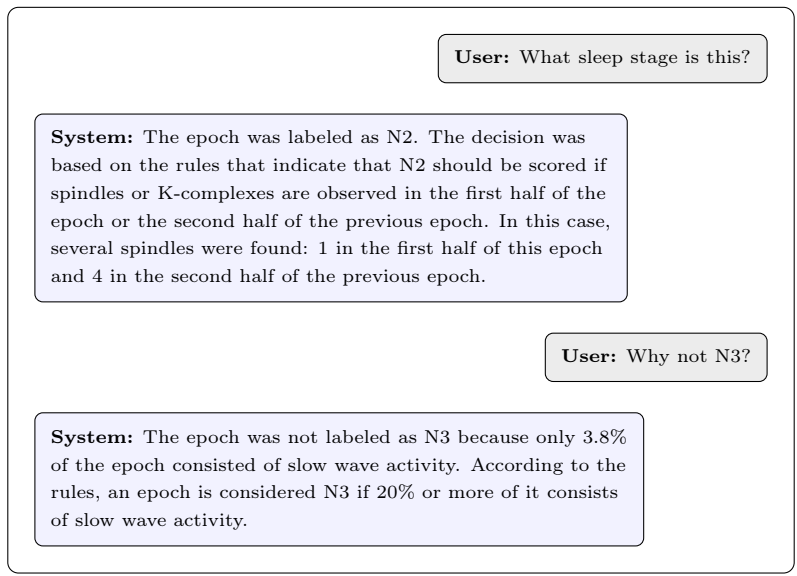
\begin{figure}[t]
\centering
\begin{tikzpicture}[
  font=\fontsize{7.0}{10.5}\selectfont,
  chatuser/.style={
    draw,
    rounded corners=3pt,
    fill=gray!15,
    inner sep=6pt
  },
  chatllm/.style={
    draw,
    rounded corners=3pt,
    fill=blue!5,
    inner sep=6pt
  },
  outerbox/.style={
    draw,
    rounded corners=4pt,
    inner sep=10pt
  }
]

\coordinate (L) at (0,0);
\coordinate (R) at (0.80\linewidth,0);

\node[chatuser, anchor=north east] (u1) at (R)
{\begin{varwidth}{0.74\linewidth}\raggedright
\textbf{User:} What sleep stage is this?
\end{varwidth}};

\node[chatllm, anchor=north west] (a1)
at ([yshift=-4mm]u1.south west -| L)
{\begin{varwidth}{0.64\linewidth}\raggedright
\textbf{System:} The epoch was labeled as N2. The decision was based on the
rules that indicate that N2 should be scored if spindles or K-complexes are
observed in the first half of the epoch or the second half of the previous
epoch. In this case, several spindles were found: 1 in the first half of this
epoch and 4 in the second half of the previous epoch.
\end{varwidth}};

\node[chatuser, anchor=north east] (u2)
at ([yshift=-4mm]a1.south east -| R)
{\begin{varwidth}{0.74\linewidth}\raggedright
\textbf{User:} Why not N3?
\end{varwidth}};

\node[chatllm, anchor=north west] (a2)
at ([yshift=-4mm]u2.south west -| L)
{\begin{varwidth}{0.64\linewidth}\raggedright
\textbf{System:} The epoch was not labeled as N3 because only 3.8\% of the
epoch consisted of slow wave activity. According to the rules, an epoch is
considered N3 if 20\% or more of it consists of slow wave activity.
\end{varwidth}};

\node[outerbox, fit=(u1)(a1)(u2)(a2)] {};

\end{tikzpicture}
\caption{Example of a natural-language explanation dialogue produced by the
rule-based sleep staging method, showing both the justification for the
assigned sleep stage and the rejection of an alternative candidate sleep stage. Abbreviations: N2, non-rapid eye movement sleep stage 2; N3, non-rapid eye movement sleep stage 3.}
\label{fig:chat_explanation_example}
\end{figure}

Interactive explanations can be explored in a conversation with an LLM through a chat interface. The LLM (OpenAI GPT 4) summarizes each epoch's explanation log in concise natural language and responds to the user's questions. Figure \ref{fig:chat_explanation_example} shows an example of an interactive explanation.

Given identical inputs, the method is deterministic: repeated runs produce identical hypnograms, micro-annotations and explanation logs. However, the LLM responses in the interactive explanations vary across runs, due to probabilistic sampling during the LLM's sampling process \cite{holtzman_curious_2020}.

The average runtime for a full-night recording was approximately 90 minutes on an Intel Core i7-8650U CPU, with the overwhelming majority of computation time spent in preprocessing and annotation.

\subsection{Comparison to human consensus}

To evaluate agreement with human scoring, we compared the algorithm's epoch-by-epoch sleep stage labels to the majority-vote consensus hypnogram of the ten human scorers in the dataset. 
As described in Section \ref{sec:dataset}, one PSG recording was used during the development of the algorithm to iterate on rule implementation and parameter tuning.
This recording was excluded from the test set, and we report results on this \textit{development recording} separately.

Tables \ref{tab:cm_percentages_dev} and \ref{tab:cm_counts_dev} show sleep stage confusion matrices for the development PSG, expressed as row-normalized percentages and total counts, respectively.
Tables \ref{tab:cm_percentages} and \ref{tab:cm_counts} show the same, but aggregated across all subjects in the test set. 

Overall agreement with majority-vote consensus across the dataset was 60.5\% with Cohen's $\kappa$=0.42. Agreement was highest for N2 (recall 83.5\%) and moderate for sleep stage R (recall 68.7\%) and N3 (recall 50.5\%). Wake and N1 recall were low (14.5\% and 5.9\%, respectively).

\begin{table}[t]
\caption{Confusion matrix for the single development polysomnography (excluded from the primary evaluation), comparing epoch-by-epoch predictions against the majority-vote consensus of ten human sleep experts, expressed in percentages. Each row is normalized to represent 100\% of the epochs for that specific reference sleep stage. The rows show the reference sleep stages, while the columns show the sleep stages assigned by the method. Abbreviations: N1, non-rapid eye movement sleep stage 1; N2, non-rapid eye movement sleep stage 2; N3, non-rapid eye movement sleep stage 3; R, rapid eye movement sleep.}
\label{tab:cm_percentages_dev}
\begin{tabular*}{\linewidth}{@{\extracolsep\fill}lccccc@{\extracolsep\fill}}%
\toprule
 & \multicolumn{5}{c}{\textbf{Method}} \\
\cmidrule(lr){2-6}
\textbf{Reference} & \textbf{Wake} & \textbf{N1} & \textbf{N2} & \textbf{N3} & \textbf{R} \\
\midrule
Wake & 40.7\% & 3.7\% & 37.0\% & 1.9\% & 16.7\% \\
N1 & 2.8\% & 11.1\% & 63.9\% & 0.0\% & 22.2\% \\
N2 & 0.0\% & 2.0\% & 94.8\% & 0.2\% & 2.9\% \\
N3 & 0.0\% & 10.1\% & 43.5\% & 44.2\% & 2.2\% \\
R & 0.8\% & 3.0\% & 11.4\% & 0.0\% & 84.8\% \\
\bottomrule
\end{tabular*}
\end{table}

\begin{table}[t]
\caption{Confusion matrix for the single development polysomnography (excluded from the primary evaluation), expressed in total epoch counts. The rows show the reference sleep stages, while the columns show the sleep stages assigned by the method. Abbreviations: N1, non-rapid eye movement sleep stage 1; N2, non-rapid eye movement sleep stage 2; N3, non-rapid eye movement sleep stage 3; R, rapid eye movement sleep.}
\label{tab:cm_counts_dev}
\begin{tabular*}{\linewidth}{@{\extracolsep\fill}lccccc@{\extracolsep\fill}}%
\toprule
 & \multicolumn{5}{c}{\textbf{Method}} \\
\cmidrule(lr){2-6}
\textbf{Reference} & \textbf{Wake} & \textbf{N1} & \textbf{N2} & \textbf{N3} & \textbf{R} \\
\midrule
Wake & 22 & 2 & 20 & 1 & 9 \\
N1 & 1 & 4 & 23 & 0 & 8 \\
N2 & 0 & 9 & 421 & 1 & 13 \\
N3 & 0 & 14 & 60 & 61 & 3 \\
R & 1 & 4 & 15 & 0 & 112 \\
\bottomrule
\end{tabular*}
\end{table}

\begin{table}[t]
\caption{Confusion matrix for all test recordings in the dataset, comparing the algorithm's epoch-by-epoch predictions against the majority-vote consensus of ten human sleep experts, expressed in percentages. Each row is normalized to represent 100\% of the epochs for that specific reference sleep stage. The rows show the reference sleep stages, while the columns show the sleep stages assigned by the algorithm. Abbreviations: N1, non-rapid eye movement sleep stage 1; N2, non-rapid eye movement sleep stage 2; N3, non-rapid eye movement sleep stage 3; R, rapid eye movement sleep.}
\label{tab:cm_percentages}
\begin{tabular*}{\linewidth}{@{\extracolsep\fill}lccccc@{\extracolsep\fill}}%
\toprule
 & \multicolumn{5}{c}{\textbf{Method}} \\
\cmidrule(lr){2-6}
\textbf{Reference} & \textbf{Wake (\%)} & \textbf{N1 (\%)} & \textbf{N2 (\%)} & \textbf{N3 (\%)} & \textbf{R (\%)} \\
\midrule
Wake & 14.5 & 3.7 & 43.4 & 3.4 & 35.0 \\
N1   & 0.1  & 5.9 & 54.7 & 0.3 & 39.1 \\
N2   & 0.0  & 2.7 & 83.5 & 1.7 & 12.1 \\
N3   & 0.0  & 9.2 & 24.5 & 50.5 & 15.8 \\
R    & 0.4  & 2.0 & 28.9 & 0.1 & 68.7 \\
\bottomrule
\end{tabular*}
\end{table}

\begin{table}[t]
\caption{Confusion matrix for all test recordings in the dataset, comparing the algorithm's epoch-by-epoch predictions against the majority-vote consensus of ten human sleep experts, expressed in total epoch counts. The rows show the reference sleep stages, while the columns show the sleep stages assigned by the algorithm. Abbreviations: N1, non-rapid eye movement sleep stage 1; N2, non-rapid eye movement sleep stage 2; N3, non-rapid eye movement sleep stage 3; R, rapid eye movement sleep.}
\label{tab:cm_counts}
\begin{tabular*}{\linewidth}{@{\extracolsep\fill}lccccc@{\extracolsep\fill}}%
\toprule
 & \multicolumn{5}{c}{\textbf{Method}} \\
\cmidrule(lr){2-6}
\textbf{Reference} & \textbf{Wake} & \textbf{N1} & \textbf{N2} & \textbf{N3} & \textbf{R} \\
\midrule
Wake & 895 & 228 & 2676 & 209 & 2154 \\
N1   & 2   & 191 & 1780 & 10  & 1272 \\
N2   & 7   & 553 & 17228 & 358 & 2497 \\
N3   & 0   & 805 & 2151 & 4429 & 1381 \\
R    & 36  & 181 & 2646 & 11 & 6294 \\
\bottomrule
\end{tabular*}
\end{table}

Table \ref{tab:sleep_metrics} shows the method's mean absolute difference from sleep architecture metrics derived from the majority-vote consensus hypnograms. For comparison, the table also shows the mean absolute difference between individual human scorers and the consensus of the other nine scorers. Results are shown separately for the test set and for the development recording. In the test set, the method showed larger differences from the consensus-derived metrics than individual human scorers across all evaluated sleep metrics in the test set.

\begin{table}[t]
\caption{Comparison of method and human mean absolute differences (MAD) from consensus-derived sleep metrics. \textit{Method MAD} and \textit{Human MAD} refer to the test set. \textit{Dev. Method MAD} and \textit{Dev. Human MAD} refer to the single development polysomnography excluded from the primary test-set evaluation.}
\label{tab:sleep_metrics}
\small
\setlength{\tabcolsep}{4pt}
\begin{tabular*}{\linewidth}{@{\extracolsep\fill}lcccc@{\extracolsep\fill}}
\toprule
\textbf{Metric} & \textbf{Method} & \textbf{Human} & \textbf{Dev. Method} & \textbf{Dev. Human} \\
                & \textbf{MAD}    & \textbf{MAD}   & \textbf{MAD}         & \textbf{MAD} \\
\midrule
Total Sleep Time (min) & 53.3 & 12.2 & 15.0 & 3.8 \\
Sleep Efficiency (\%) & 10.6 & 2.4 & 3.7 & 0.9 \\
Wake After Sleep Onset (min) & 53.3 & 12.2 & 15.0 & 3.8 \\
N1 Duration (min) & 19.4 & 14.9 & 1.5 & 7.8 \\
N2 Duration (min) & 64.7 & 27.9 & 47.5 & 24.9 \\
N3 Duration (min) & 40.5 & 22.6 & 37.5 & 22.4 \\
REM Duration (min) & 54.1 & 9.5 & 6.5 & 5.5 \\
N1 proportion (\%) & 4.6 & 3.5 & 0.6 & 2.1 \\
N2 proportion (\%) & 10.1 & 6.5 & 9.9 & 6.6 \\
N3 proportion (\%) & 10.6 & 5.2 & 10.3 & 5.8 \\
REM proportion (\%) & 10.0 & 2.4 & 1.0 & 1.4 \\
\bottomrule
\end{tabular*}
\end{table}

\subsection{Assessment of discrepancies}

\begin{figure*}[t]
\centerline{\includegraphics[width=\textwidth]{{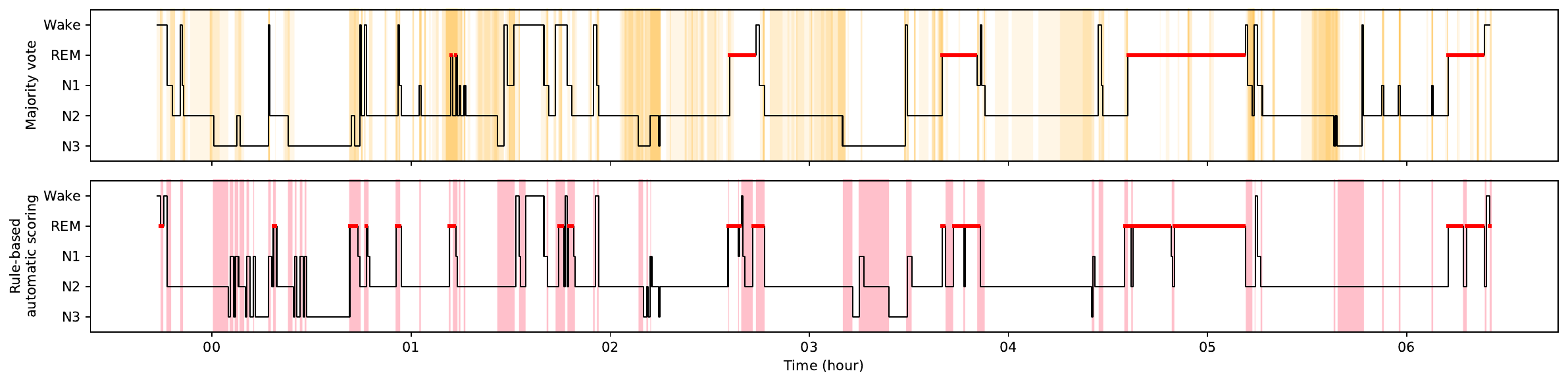}}}
\caption{The top panel shows the human consensus hypnogram for the polysomnography used during the development of the method, with areas of high human disagreement highlighted. The bottom panel shows a predicted hypnogram for the same polysomnography, with disagreements with the human consensus hypnogram highlighted in red. Abbreviations: N1, non-rapid eye movement sleep stage 1; N2, non-rapid eye movement sleep stage 2; N3, non-rapid eye movement sleep stage 3; REM, rapid eye movement sleep.}
\label{fig:hypnograms_dev}
\end{figure*}

\begin{figure*}[t]
\centerline{\includegraphics[width=\textwidth]{{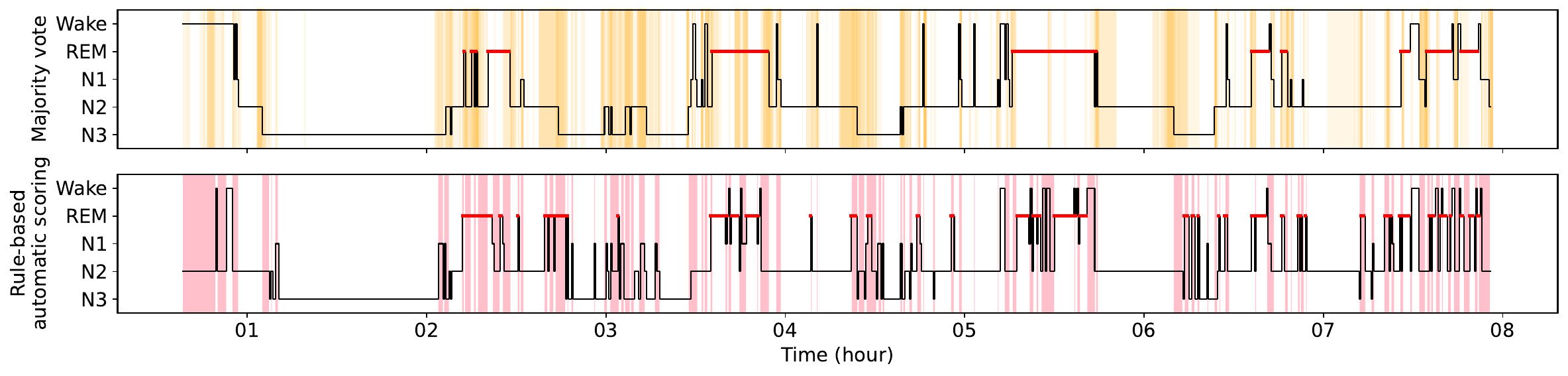}}}
\caption{The top panel shows the human consensus hypnogram for one of the test polysomnographies with areas of high human disagreement highlighted. The bottom panel shows a predicted hypnogram for the same polysomnography, with disagreements with the human consensus hypnogram highlighted in red. Abbreviations: N1, non-rapid eye movement sleep stage 1; N2, non-rapid eye movement sleep stage 2; N3, non-rapid eye movement sleep stage 3; REM, rapid eye movement sleep.}
\label{fig:hypnograms_test}
\end{figure*}

\begin{figure}[b]
\centerline{\includegraphics[width=0.75\columnwidth]{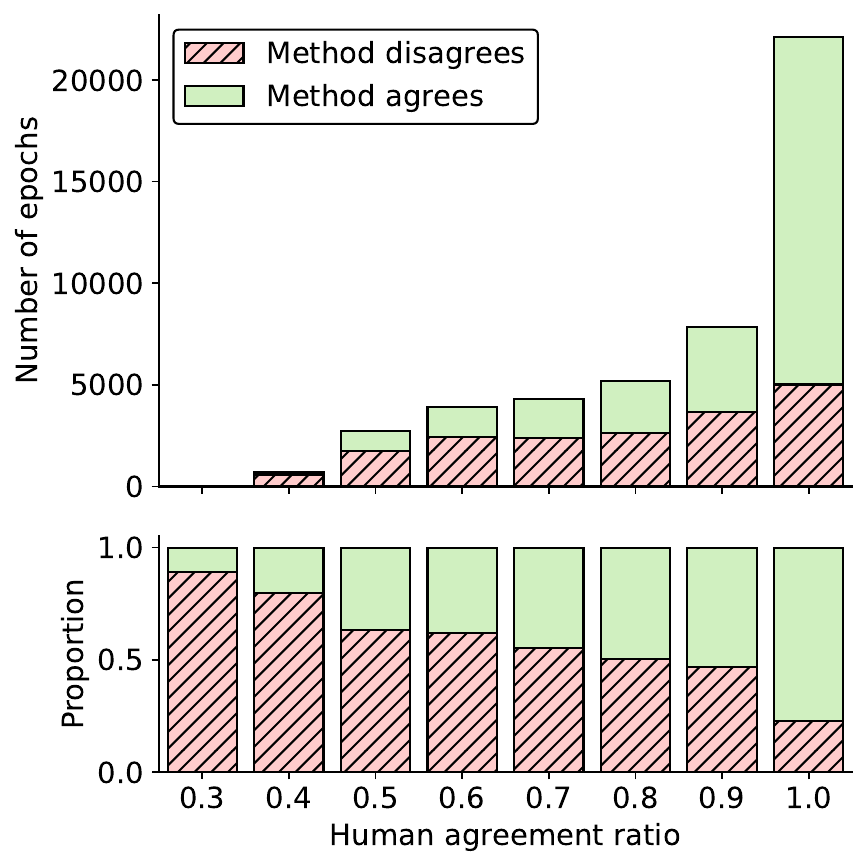}}
\caption{Distribution of human inter-scorer agreement for epochs where the rule-based algorithm agreed versus disagreed with the 10-scorer majority-vote reference. The horizontal axis shows the per-epoch agreement ratio (fraction of scorers assigning the modal sleep stage). Disagreements with the reference are proportionally more common where human agreement is lower.}
\label{fig:agreement_distribution}
\end{figure}

Figures \ref{fig:hypnograms_dev} and \ref{fig:hypnograms_test} show examples of hypnograms comparing the algorithm's sleep stage predictions to the majority-vote consensus. In both figures, epochs where the algorithm disagrees with the consensus sleep stage are highlighted in red. The consensus hypnogram is highlighted in a progressively darker shade of yellow, depending on the level of disagreement among the 10 scorers. Many epochs where the algorithm disagreed with the reference occurred in areas where human agreement was low. This can also be seen in Figure \ref{fig:agreement_distribution}, which shows the distribution of human agreement ratios for epochs where the algorithm agreed with the reference, compared with epochs where it disagreed. 

The reviewer's discrepancy assessment suggested that when the reviewer agreed with the method's sleep stage label, the decision was usually supported by the appropriate reasoning. In 71\% of epochs where the reviewer agreed with the method's sleep stage label, the micro-annotations, rule application, and definite/transition assignment were all deemed correct by the reviewer.
In epochs where the reviewer disagreed with the method's label, the micro-annotations were judged incorrect in all cases.
These findings suggest that, among the reviewed examples, reviewer-method disagreements were primarily associated with micro-annotation quality, rather than the rule logic itself.

\section{Discussion}
\label{sec:discussion}

This work proposes a rule-based automatic sleep staging method that operationalizes the AASM scoring manual \citep{troester_aasm_2023} as executable code, giving decisions that are fully deterministic and explainable \citep{stanley_future_2023,hulsen_explainable_2023}. 
Rather than treating sleep staging solely as a statistical prediction problem, the method codifies the clinical decision process, such that the reasoning can be inspected and rendered as static or interactive natural-language explanations.
In addition to the sleep stage classifier, the method thus provides a tool for auditing, education, and explanation in clinical settings where alignment with standardized scoring rules is of importance.

The method clearly does not outperform contemporary ML and DL methods in terms of scoring accuracy \citep{alsolai_systematic_2022,fiorillo_automated_2019,gaiduk_current_2023,fiorillo_u-sleeps_2023,phan_l-seqsleepnet_2023}.
Our results show that although recall for sleep stages N2 and R is strong, the method showed lower agreement with the reference for Wake, N1, and, to a lesser extent, N3.
However, it provides an alternative, complementary way to estimate sleep stages.
This opens up the possibility of hybrid automatic sleep staging methods \citep{liang_rule-based_2012,gunnarsdottir_novel_2020}, where a deterministic and auditable method, grounded in clinical rules, runs in parallel with ML/DL sleep staging methods, as a reference or cross-check.

The reviewer-based discrepancy assessment performed in this work clarifies where the current implementation differed from the majority-vote reference and how those differences arose.
Many disagreements between the algorithm and the human consensus scoring occurred in epochs where human inter-scorer agreement was also low, suggesting that disagreements often occur in ambiguous regions of sleep.

The assessment also showed that when the reviewer disagreed with the algorithm's sleep stage label, the micro-annotations were judged to be incorrect, pointing to the micro-annotation quality as the main technical limitation of the current implementation.

The micro-annotations were not independently validated against expert event-level annotations. In particular, alpha rhythm and LAMF were operationalized using PSD-based heuristics rather than a separately validated detector.
Future work should therefore validate and improve the available micro-annotations, including alpha rhythm, LAMF, REMs, spindles, SWA, chin EMG tone, and missing detectors such as K-complex detection.
Wake classification was particularly difficult to operationalize. This may reflect that some visual cues used by human scorers to recognize Wake epochs are less straightforward to translate into computational rules, especially in noisy epochs with movement artifacts.
The scoring manual is written to support expert human judgment but leaves ambiguity for implementation \citep{stanley_future_2023}.
For example, some scoring criteria depend on judgment of whether the majority of an epoch contains evidence for a sleep stage, without specifying how that evidence should be segmented or aggregated computationally \citep{troester_aasm_2023}.
Future iterations of scoring manuals could benefit from more operationally precise phrasing, fit for algorithmic implementation.

\section{Conclusion}

We presented a deterministic, rule-based sleep staging method that operationalizes the AASM scoring manual as executable code and records the reasoning behind each epoch-by-epoch decision as an inspectable explanation trace. The method was evaluated on 50 full-night type II PSG recordings scored independently by 10 human sleep experts, using the majority-vote consensus as reference. Overall agreement with the consensus was moderate (60.5\%, $\kappa$=0.42), with strong recall for sleep stages N2 and R, and lower recall for Wake, N1, and N3. A reviewer-based discrepancy assessment indicated that disagreements were primarily associated with the quality of micro-annotations rather than the rule logic itself.

Taken together, these findings suggest that even if rule-based sleep staging is not yet competitive with DL in terms of agreement with human-scored references, it remains valuable as a transparent complement to black-box methods.
Its current limitations appear in the lack of reliable micro-annotation and the ambiguity of a scoring manual written for human experts.

\section*{Author contributions}

EH conceived the study, developed and implemented the method, and led the manuscript writing.
SS contributed to the conception and testing of the method and manuscript writing.
KP contributed to data analysis, interpretation, and manuscript writing.
ASI, ESA, and MÓ contributed to study design, acquisition of funding, interpretation of results, and manuscript writing.
All authors reviewed and approved the final manuscript.

\section*{Acknowledgments}
We thank Kristín Anna Ólafsdóttir, Heiður Grétarsdóttir, and Gabriel Jouan for their help during the development of the method, the 10 scorers of the data set, as well as other staff and students of the Sleep Revolution project who contributed to data collection and analysis.
LLMs were used during the writing of this paper to enhance the readability of the text.

\section*{Financial disclosure}

This project was funded by the European Union’s 2020 Research and Innovation Program under Grant 965417 and the Icelandic Research Fund under Doctoral Student Grant 2410607-051.

\section*{Conflict of interest}

ESA discloses lecture fees from Nox Medical, ResMed, Jazz Pharmaceuticals, Linde Healthcare, Wink Sleep, Apnimed, and Vistor. ESA is a previous member of the Philips Sleep Medicine \& Innovation Medical Advisory Board and the Lille Medical Advisory Board and is currently on the Sleep Cycle Data Monitoring Committee. The other authors declare no potential conflict of interest.

\bibliographystyle{plainnat}   
\bibliography{references}      

\end{document}